# Supersymmetric Laser Arrays

Mohammad P. Hokmabadi[1], Nicholas S. Nye[1], Ramy El-Ganainy[2], Demetrios N. Christodoulides[1], Mercedeh Khajavikhan[1*]

[1]College of Optics & Photonics-CREOL, University of Central Florida, Orlando, Florida 32816, USA

[2]Department of Physics and Henes Center for Quantum Phenomena, Michigan Technological University, Houghton, Michigan 49931, USA

[*]Corresponding author: mercedeh@creol.ucf.edu

**Abstract: The theoretical framework of supersymmetry (SUSY) aims to relate bosons and fermions- two profoundly different species of particles- and their interactions. While this space-time symmetry is seen to provide an elegant solution to many unanswered questions in high-energy physics, its experimental verification has so far remained elusive. Here, we demonstrate that, notions from supersymmetry can be strategically utilized in optics in order to address one of the longstanding challenges in laser science. In this regard, a supersymmetric laser array is realized, capable of emitting exclusively in its fundamental transverse mode. Our results not only pave the way towards devising new schemes for scaling up radiance in integrated lasers, but also on a more fundamental level, they could shed light on the intriguing synergy between non-Hermiticity and supersymmetry.**

Symmetries play a fundamental role in physical sciences. Symmetry principles ensure energy and momentum conservation and dictate the allowable dynamical laws governing our world. The Lorentz invariance embodied in Maxwell's equations was crucial in developing the theory of relativity, while the exchange symmetry allows one to classify fundamental particles as either bosons or fermions. In high-energy physics, other overarching symmetries like that of charge-parity-time (CPT) and supersymmetry (SUSY) have also emerged as a means to unveil the laws of nature [1,2]. SUSY, first proposed within the context of particle physics as an extension of the Poincare space-time symmetry, makes an ambitious attempt to provide a unified description of all fundamental interactions. In general, SUSY relates bosonic and fermionic degrees of freedom in a cohesive fashion. This directly implies that each type of boson has a supersymmetric counterpart, a superpartner fermion, and vice versa [3]. Even though the full ramification of SUSY in high energy physics is still a matter of debate that awaits experimental validation, supersymmetric techniques have already found their way into low energy physics, condensed matter, statistical mechanic, nonlinear dynamics and soliton theory as well as in stochastic processes and BCS-type theories, to mention a few [4-9].

Shortly after the discovery of semiconductor lasers, it was proposed that integrated arrays of such emitters may provide a viable avenue in scaling up the radiance (power per unit area per unit solid angle), without running into complications arising from nonlinearities and filamentation in broad area devices [10]. Unfortunately, however, such arrays tend to support multiple spatial modes (supermodes), an undesirable behavior that in turn degrades the quality of the emitted beam. This has since fueled a flurry of activities in search of strategies that enable the generation of high power and diffraction-limited coherent beams by enforcing the coupled laser array to operate in the fundamental (in-phase) mode. In this regard, several schemes have been developed, using for example resonant leaky-wave coupling in antiguided arrangements [11], Talbot effects [12], common antenna feedback [13], to name a few [14-16]. Of interest will be to devise fully integrated global approaches that apply to any type of active arrays in order to enforce single-mode lasing in the fundamental transverse supermode. To address this issue, here we report the first realization of a supersymmetric laser array. This lattice emits in its fundamental mode in a stable fashion, as evidenced from far-field and spectral measurements. In this SUSY arrangement, the main array is paired with a lossy superpartner, whose role is to suppress all undesired higher-order modes while at the same time enhancing the gain seen by the fundamental supermode of the primary lattice. In implementing such lasers, we made use of the SUSY formalism first proposed by Witten [17].

Within the context of non-relativistic quantum mechanics, supersymmetric isospectrality can be established provided that the Hamiltonian of the system, $H^{(1)}$, is factorized in terms of two operators $A$ and $A^\dagger$, i.e. $H^{(1)} = A^\dagger A$ [5]. Similarly, a superpartner Hamiltonian $H^{(2)}$ can be constructed via $H^{(2)} = A A^\dagger$ by exchanging the action of these two operators. If one now assumes that $|\varphi\rangle^{(1)}$ represents an eigenstate of $H^{(1)}$ with an eigenvalue $\lambda^{(1)}$, i.e. $H^{(1)}|\varphi\rangle^{(1)} = A^\dagger A|\varphi\rangle^{(1)} = \lambda^{(1)} |\varphi\rangle^{(1)}$, then it follows that $AH^{(1)}|\varphi\rangle^{(1)} = (A A^\dagger)A|\varphi\rangle^{(1)} = H^{(2)}A|\varphi\rangle^{(1)} = \lambda^{(1)} A |\varphi\rangle^{(1)}$. Hence, $A|\varphi\rangle^{(1)}$ is an eigenvector of $H^{(2)}$ with an eigenvalue $\lambda^{(1)}$. This immediately indicates that the two Hamiltonians are isospectral since they exhibit identical eigenenergies, i.e. $\lambda^{(2)} = \lambda^{(1)}$, while their eigenstates can be pairwise converted into one another through the action of the $A, A^\dagger$ operators: $|\varphi\rangle^{(2)} = A |\varphi\rangle^{(1)}$ and $|\varphi\rangle^{(1)} = A^\dagger|\varphi\rangle^{(2)}$. If the ground state of $H^{(1)}$ is annihilated by the action of the operator $A$, then the eigenenergy associated with the ground state of $H^{(1)}$ is zero, and therefore it will not have a corresponding state in $H^{(2)}$. In other words, all the eigenvalues associated with the states of $H^{(1)}$ and $H^{(2)}$ are exactly matched except for the lowest energy state of $H^{(1)}$. When this is the case, then the SUSY is called unbroken. Otherwise, if the ground state of $H^{(1)}$ has a counterpart in its superpartner $H^{(2)}$, with the same eigenvalue, the supersymmetry is broken. Using this approach in 1D Schrödinger problems, one can always identify two SUSY potential functions, $V^{(1)}(x)$ and the superpartner $V^{(2)}(x)$, that can be entirely isospectral except for the lowest energy state of $V^{(1)}$ (Fig. 1A) [5]. In optics, SUSY can be introduced by exploiting the mathematical isomorphism between the Schrödinger and the optical wave equation [18]. In this setting, the optical refractive index profile plays the role of the potential $V(x)$, which in the context of supersymmetry can be used for mode conversion [19,20], transformation optics [21], design of Bragg gratings [22], and Bloch-like waves in random-walk potentials [23], to mention a few [24- 26]. However, the implications of SUSY isospectrality in active platforms, as well as its interplay with nonlinearity and non-Hermiticity has so far remained unexplored. Here we lay the groundwork for such studies by demonstrating a SUSY-based laser.

Figure 1B depicts a schematic of the proposed supersymmetric laser. The primary array in this SUSY arrangement is synthesized by coupling five identical ridge-waveguide cavities of length $L$. The individual waveguide elements are designed to support only the lowest order transverse mode ($TE_0$). Consequently, each element on its own, is expected to support resonances at angular frequencies $\omega_m = m\pi c/(Ln)$, where $n$ represents the effective index associated with the $TE_0$ mode, $c$ is the speed of light in vacuum, and $m$ is an integer representing the longitudinal mode index. The evanescent coupling between the five waveguides, causes every such resonant frequency, $\omega_m$, to split into a cluster of five frequencies, corresponding to the five supermodes of the active array. Optical supersymmetric strategies are then employed to build a superpartner index profile that has propagation eigenvalues that match those of the four higher-order supermodes associated with the main (primary) array [27].

The SUSY laser arrays were realized on an InP wafer with InGaAsP quantum wells as the gain material. In doing so, we used electron beam lithography and plasma etching techniques to define the structures (Supplementary Information, Part I). Figure 2A displays a scanning electron microscopy (SEM) image of the fabricated SUSY structures. Finite element method (FEM) simulations were performed to determine the modal content of these structures. Details on the design and simulation of these arrangements can be found in Supplementary Information, Part II. Figure 2B depicts the intensity profiles of all the modes supported by this SUSY configuration. The performance of the SUSY laser was then assessed by means of a custom-made optical setup (Supplementary Information, Part III). The arrays are optically pumped at a wavelength of 1064 nm, emitted from a fiber laser. Spatial masks are deployed to selectively pump different regions. The coherent radiation (centered around 1450 nm) emerging from the cleaved facets of the lasers is monitored by both a spectrometer and an infrared camera, after blocking remnants of the pump emission by means of a notch filter. The diffraction angles associated with the far-field emissions along the laser's slow axis, were determined by raster scanning a rectangular aperture placed in front of the array. In every step, the total emitted power from the laser was also measured by a photodiode.

The spectral response, far-field emission, and light-light characteristics are compared for three different configurations: (i) a single ridge waveguide lasing element, (ii) a standard laser array involving five evanescently coupled ridge cavities, and (iii) a SUSY laser array comprised of a primary active five-element lattice and its corresponding four-element lossy superpartner. In the latter configuration, the two lattices are fabricated in a close proximity to each other and are therefore coupled. The system is judiciously designed so as SUSY is unbroken. This was achieved by appropriately varying the widths and spacings (equivalently the corresponding effective refractive indices) of the ridge elements in the superpartner array. Schematic representations of these three lasers are provided in the insets of Figs. 3A, C, and E, respectively. The lasers are uniformly pumped at an average power density level that is approximately 4 times the threshold of the SUSY laser. Loss is introduced in the superpartner array by blocking the pump beam using a knife-edge. Under these pumping conditions, the single element cavity lases in a few longitudinal modes (in the $TE_0$ mode), at wavelengths around 1443 nm (Fig. 3A). When the five-element standard laser array is exposed to the same pump power density, we found that each longitudinal mode now splits into five lines corresponding to the resonances of the five supermodes involved (Fig. 3C). This multimode operation can to lead to a significant deterioration of the beam quality emitted by such a lattice. In contrast, when the SUSY laser array is illuminated at the same pump intensity level (while the superpartner is blocked), the device emits in a single transverse supermode (Fig. 3E). Moreover, the peak intensity produced by this SUSY laser is now 4.2 times higher than that from the standard laser array (i.e. without superpartner), and 8.5 times larger than that from the single element laser. These results clearly indicate that in a SUSY laser arrangement, all higher-order transverse modes are indeed suppressed in favor of the fundamental mode.

To further verify the anticipated SUSY response, the far-field radiation from these three laser systems was collected along with the diffraction profiles in the direction of the slow axis (parallel to the wafer). These measurements are correspondingly displayed in Figs. 3B, D, and F. A comparison between these three radiation patterns reveals a striking difference in the way a SUSY laser operates. As opposed to the standard laser array, whose far-field exhibits a multi-lobe profile with a diffraction angle of $\sim 19°$ (Fig. 3D), the far-field of the SUSY array displays a single bright spot having instead a much smaller divergence angle of $\sim 5.8°$ (Fig. 3F). This low divergence behavior is a characteristic attribute of a laser array operating only in its in-phase lowest-order mode [28]. In addition, in the standard array system, we observe a multi-lobe far-field pattern that changes with pump intensity (Supplementary Information, Part IV). Even more importantly, the beam spot size associated with the SUSY laser is narrower than that of a single laser element ($\sim 12°$) as shown in Fig. 3B- indicating a higher brightness associated with the SUSY arrangement. These experimentally obtained diffraction patterns are in good agreement with numerical simulations (Supplementary Information, Part V).

Finally, the light-light curves corresponding to these three lasers and the evolution of their spectra are depicted in Fig. 4A & B, respectively. As expected, both SUSY and standard laser arrays outperform the single element laser in terms of output power (Figure 4A). When the overall output power is compared, the two arrays (standard and SUSY) were found to exhibit similar thresholds and slope efficiencies. On the other hand, Fig. 4B provides valuable information as to the lasing onset for higher-order supermodes. More specifically, as the pump power is gradually increased above the threshold, the higher-order modes of the standard laser array start to successively emerge in the spectrum (blue lines in Fig. 4B), while the SUSY array still lases in its fundamental transverse mode with larger spectral peaks (red lines in Fig. 4B). These observations further confirm that indeed in a SUSY laser, all undesired higher-order modes are effectively eliminated via coupling to the lossy superpartner- giving the opportunity to the fundamental mode to prevail.

It also shows that the loss of the superpartner array has no effect on the efficiency of the SUSY laser. The resilience of these arrays to fabrication imperfections and errors is discussed in Supplementary Information Part VI, where a sensitivity analysis clearly shows that the SUSY transformations are robust against first-order perturbations. This feature to some degree relaxes the requirements for phase-locking in the presence of fabrication and environmental errors, as well as variations of index due to non-linearities or temperature because of intensity inhomogeneities.

In conclusion, by harnessing notions from supersymmetry, we present the first realization of an integrated supersymmetric laser array. Our results indicate that the existence of an unbroken SUSY phase in conjunction with a judicious pumping of the laser array, can promote the in-phase supermode, thus resulting to a high radiance emission. This new mechanism of phase-locking is resilient to first order deviations in fabrication, and provides a global approach that can be systematically applied to a wide range of coupled active lattices. Our results may have practical implications in designing high brightness single mode laser arrays from UV to Mid-IR sources, while introducing a unique platform to study, at a fundamental level, the interplay between non-Hermiticity and supersymmetry.

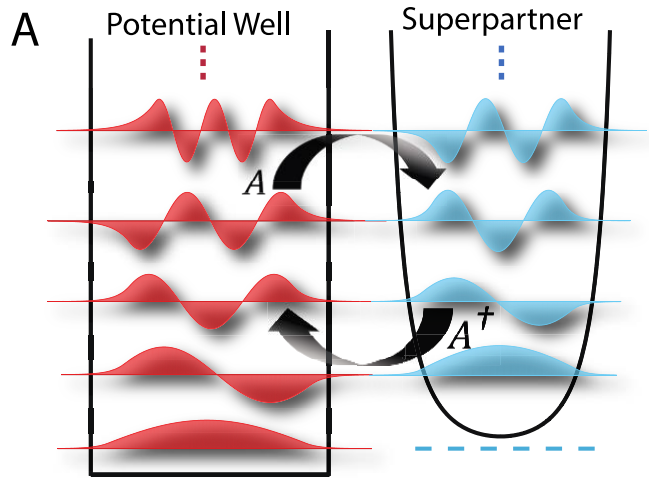

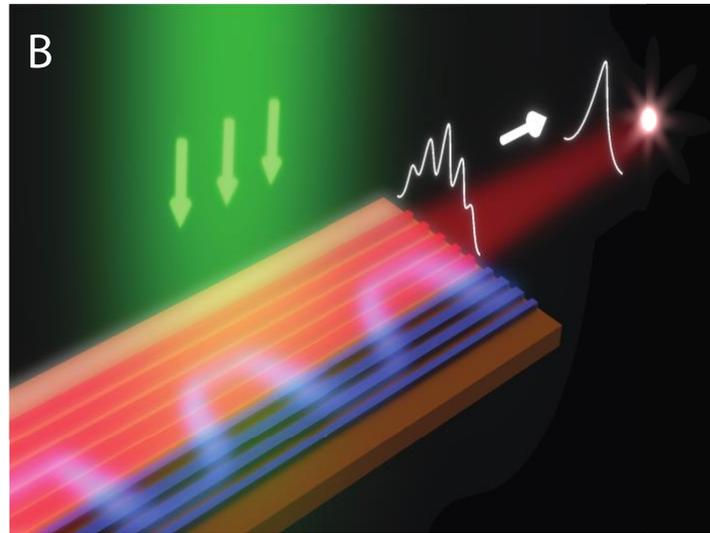

**Fig. 1. Operation principle of SUSY laser array.** (**A**) An infinite potential well and its superpartner in the unbroken SUSY regime. Apart from the ground state, all the eigenvalues of the primary potential are exactly matched to those of the superpartner. The eigenfunctions of the primary potential and its supersymmetric counterpart are transformed into one another through the action of the operators $A$ and $A^{\dagger}$. (**B**) A schematic representation of a SUSY laser array involving a primary active lattice (red) coupled to its lossy superpartner (blue). The SUSY laser emits exclusively in the fundamental in-phase mode.

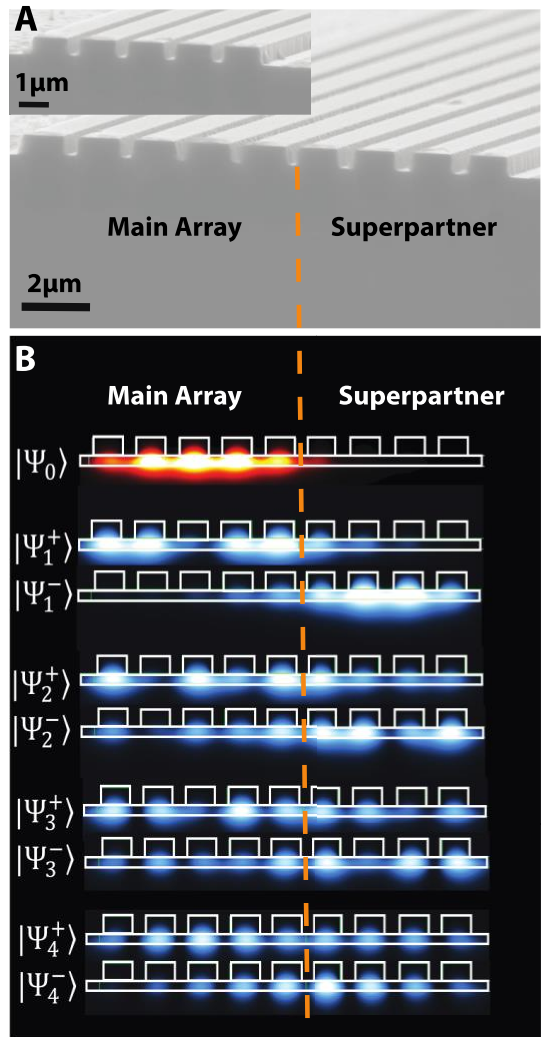

**Fig. 2. A scanning electron microscope image (SEM) and modal field profiles of the SUSY laser array.** (**A**) SEM image of a fabricated SUSY lattice comprised of a five-element primary array, positioned in close proximity ($400\,nm$) to a four-element superpartner. The inset shows a stand-alone five-element laser array. (**B**) Intensity distributions associated with the eigenmodes supported by the SUSY arrangement, as obtained from numerical simulations. The fundamental mode of the five-element laser is only confined in the main array, while all the higher-order modes are coupled to the lossy superpartner. The dashed line illustrates the boundary between the main and the superpartner structures.

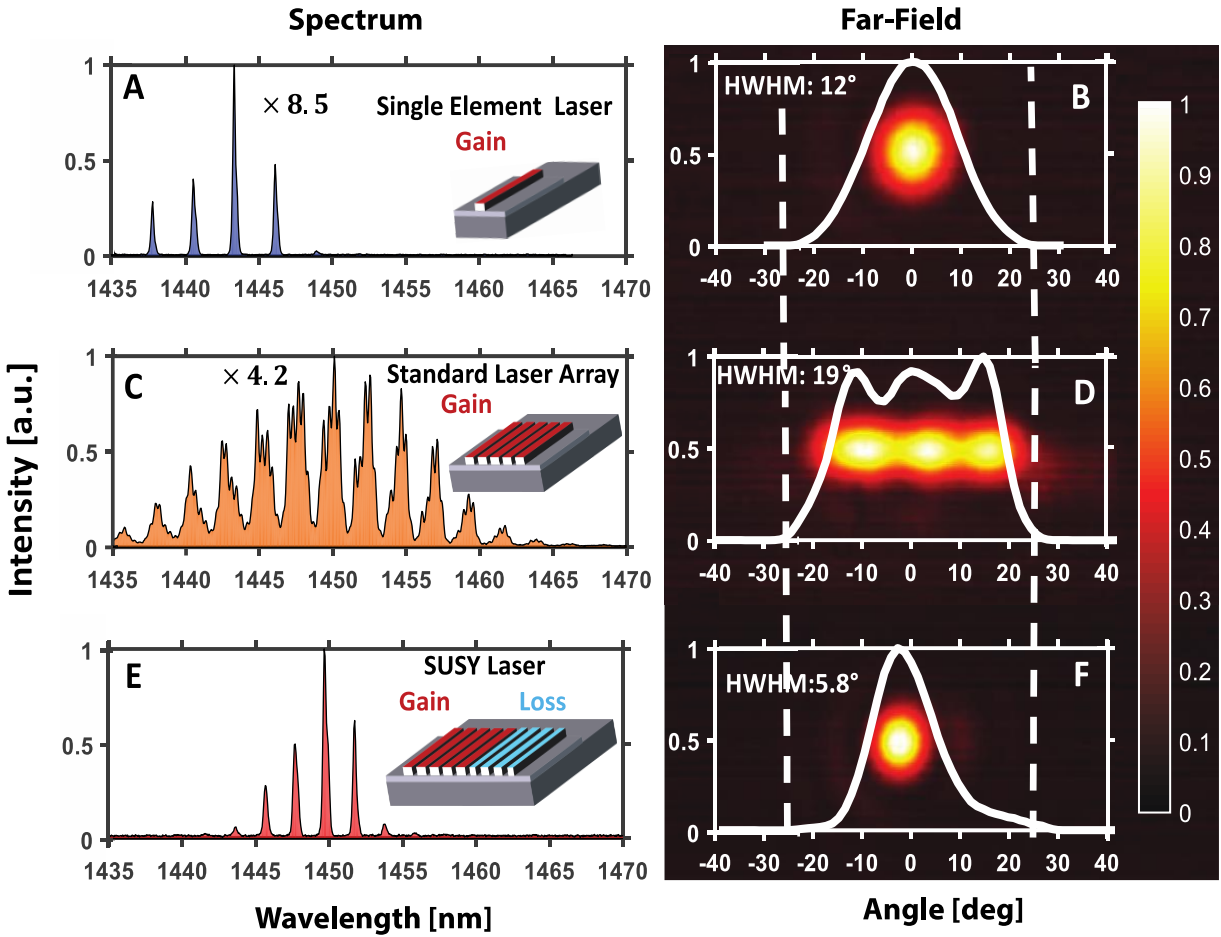

**Fig. 3. Spectral and far-field characteristics of the SUSY laser array.** Emission spectrum of a (**A**) single laser cavity, (**C**) standard five-element laser array, and (**E**) corresponding SUSY laser arrangement. The vertical axes are normalized to the spectrum of the SUSY laser. Each longitudinal resonant frequency, in the spectrum of the standard array splits into five lines- corresponding to the five transverse supermodes. In contrast, the spectrum of the SUSY array is free from such undesired resonances, indicating that all higher order modes are suppressed. (**B, D, F**) Far-field diffraction patterns from the corresponding lasers. The measured diffraction angle associated with the SUSY laser (~5.8°) is smaller than that of the standard laser array (~19°) and a single waveguide laser (~12°).

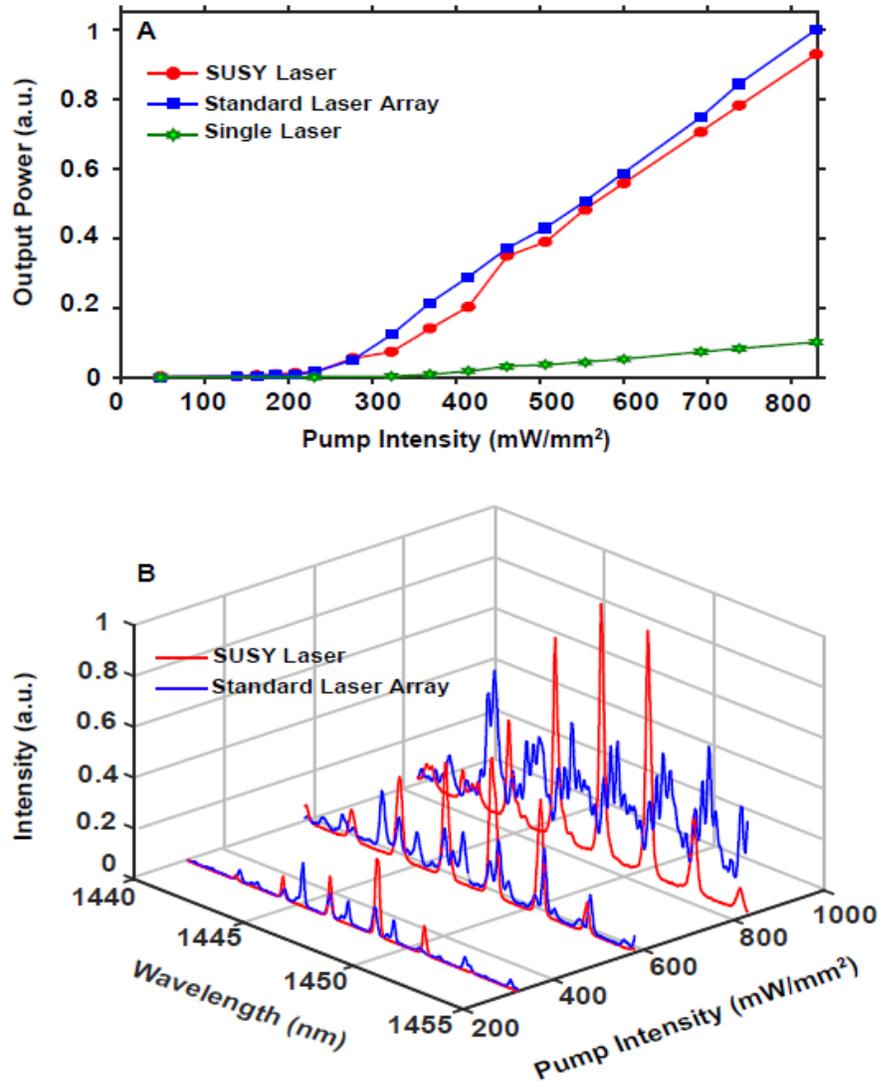

**Fig. 4. Emission characteristics.** (**A**) The light-light curves corresponding to a single laser element (green line), a five-element laser array (blue line), and the SUSY laser arrangement (red line). The output power and slope efficiency of the standard array and the SUSY laser are comparable, and exceed that from a single cavity laser. (**B**) The spectral evolution behaviors of the standard and SUSY lasers are compared. The standard laser array (blue line) is highly multimoded even at pump levels slightly above threshold, while the SUSY laser array remains transversely single-moded.

# Supersymmetric Laser Arrays

## Supplementary Information

Mohammad P. Hokmabadi[1], Nicholas S. Nye[1], Ramy El-Ganainy[2], Demetrios N. Christodoulides[1], Mercedeh Khajavikhan[1*]

[1] College of Optics & Photonics-CREOL, University of Central Florida, Orlando, Florida 32816, USA

[2] Department of Physics and Henes Center for Quantum Phenomena, Michigan Technological University, Houghton, Michigan 49931, USA

*Corresponding author: mercedeh@creol.ucf.edu

**Part I. Sample fabrication**

The steps involved in the fabrication of the SUSY laser arrays are schematically depicted in Fig. S1. The wafer under process is composed of a multiple quantum wells gain layer, grown on an intrinsic InP substrate, and covered by a 500 nm thick undoped InP film (Fig. S1A). The gain layer consists of 10 quantum wells of $In_{x=0.737}Ga_{1-x}As_{y=0.569}P_{1-y}$ (20 nm)/$In_{x=0.564}Ga_{1-x}As_{y=0.933}P_{1-y}$ (10 nm). After cleaning, hydrogen silsesquioxane (HSQ) solution was spun over the wafer to obtain a 100 nm thick layer of the negative tone inorganic electron beam resist (Fig. S1B). Laser arrays with different widths and spacings were patterned after exposure to the electron beam, and developed by tetramethylammonium hydroxide (TMAH) solution (Fig. S1C). These patterns were then transferred to the wafer through dry etching which is performed by a combination of reactive ion etching and inductively coupled plasma (RIE-ICP) processes ($H_2$:$CH_4$:AR, 40:10:20 SCCM, RIE power: 150 W, ICP power: 150 W, Pressure:35 mT). Overall, 500 nm of InP is removed in selected areas, leaving behind the ridge waveguides of the laser cavities (Fig. S1D). The wafer is cleaned in oxygen plasma in order to remove the remnant of organic contaminations and polymers, accumulated during the dry etching process. Next, the rear facets of the laser cavities were defined. This was accomplished in a four-step process. First, a 500 nm thick $SiO_2$ mask was deposited by means of plasma-enhanced chemical vapor deposition (PECVD). The ridge waveguides were partially covered by a negative tone photoresist (NR7-3000) through spinning and UV-lithography, allowing the uncovered $SiO_2$ layer to be dry etched. The rear facets of the lasers were formed by further RIE-ICP dry etching of the gain layer in places where the $SiO_2$ masks were absent (Fig. S1E). The exposed HSQ and the remaining $SiO_2$ was removed by dissolving it in a buffer oxide etchant (BOE) solution. Finally, the front facets of the lasers were defined by cleaving the wafer (Fig. S1F).

**Part II. Design procedure of the supersymmetric laser array**

For a given main laser array, the superpartner counterpart is designed by applying SUSY techniques. For the laser discussed in the main text, the primary array consists of five single element ridge waveguide cavities with widths of $1000\ nm$ and spacings of $400\ nm$ (Fig. S2A). To design the SUSY partner, the Hamiltonian of the primary lattice is discretized into an $N \times N$ tridiagonal matrix. The elements of this matrix are given by $H_{n,n}^{(1)} = \lambda_s$ and $H_{n,n+1}^{(1)} = H_{n+1,n}^{(1)} = \kappa$, where $\lambda_s$ is the eigenvalue (effective index) of the single element laser constructing the main array and $\kappa$ is the coupling constant of the two adjacent single cavities. The Hamiltonian of the superpartner array is then obtained by $H^{(2)} = (RQ + \lambda_0 I)_{(N-1)} = (Q^T H^{(1)} Q)_{(N-1)}$ in which $\lambda_0$ is the eigenenergy of the fundamental mode of the main array, $I$ is the identity matrix of dimensions $N \times N$, and $Q$ and $R$ are the $QR$ factorization matrices of $H^{(1)} - \lambda_1 I$ [1-3]. Here, the subscript $(N - 1)$ means that $H^{(2)}$ is built by only choosing the upper-left block diagonal matrix having the dimensions $(N - 1) \times (N - 1)$. Finally, the SUSY laser configuration is realized by evanescently coupling the main and the superpartner arrays. A schematic of the SUSY laser array is depicted in Fig. S2A, where the dimensions of the various parts are specified.

The simulated eigenmodes of the main array and the superpartner, before coupling to each other are depicted in Figs.S2 B & C. In these simulations, a two-dimensional finite element numerical method is employed to obtain the eigenstates supported by each array. The refractive index of InP and InGaAsP in the simulations are considered to be 3.4 and 3.14, respectively. The simulations are performed by the mode analysis module in COMSOL Multiphysics package.

**Part III. Measurement setup**

The experimental setup is schematically illustrated in Fig. S3. The pump beam (wavelength: 1064 nm, pulse duration: 9 ns, repetition rate: 290 kHz) was provided by a fiber laser. The appropriately shaped pump beam illuminates the sample after passing through a 20X objective. An image of a knife edge is used in order to block the pump beam from the superpartner array. Without the knife edge, the diameter of the beam on the sample is approximately $120\ \mu m$. The same 20X objective lens also serves to image the top of the sample, illuminated by an amplified spontaneous emission passed through a rotating diffuser, onto an infrared CCD camera. This imaging system allows one to precisely adjust the location of the pump beam (and knife edge) with respect to the array. The laser output is collected form the cleaved facet at the edge of the sample by means of a 50X objective lens. The collected emission is then directed into a spectrometer, a power-meter, as well as a CCD infrared camera.

To measure the diffraction angle associated with the emission of the lasers, a rectangular slit is placed between 50X objective lens and the sample (Fig. S4). The rectangular slit is raster-scanned by using a stepper motor parallel to the surface of the sample, i.e. in the direction of lasers' slow axis. To construct the transverse profile of the beam, at each step, the output power passing through the slit is measured by a photodiode operating in the lock-in detection scheme. The slit is then shifted to a new position along the optical axis, where the diffraction angle measurement is repeated. Since the exact distance between these two points along the lens' optical axis is known, the divergence angle of the emerging beam can be determined from these two measurements.

**Part IV. Far-field divergence angle: comparing simulation and measurement results**

To obtain the simulated far-field patterns emitted from the laser cavities, first the near-field modes supported by the corresponding ridge waveguides are numerically calculated using finite element method (FEM). Then, the far-field profiles and diffraction angles are calculated using Fourier transform techniques. Figure S5A shows the result of the far-field simulation for a SUSY laser (when only the main array is pumped). The simulation results are in good agreement with the experimental measurements (Fig. S5B). A similar numerical analysis, performed on the single element laser, also matches well with our experimental observations (Fig. S5C, D). It should be noted that, even though both SUSY and single element lasers generate diffraction limited far-field patterns, the beam from the SUSY arrangement has a considerably smaller spot size. This feature in combination with the higher power extracted from such SUSY lasers are direct outcome of radiance (brightness) scaling that is unique to phase-locked arrays.

Figures S5E &G display the calculated far-field emissions of the five-element standard laser array when several supermodes are simultaneously excited. For example, in these figures, the percentage

of power distribution among five supermodes involved (from in-phase to out-of-phase modes) are (30, 22, 8, 8, 32), and (30, 13, 19, 19, 19), respectively. These far-field patterns closely resemble the experimental far-field profiles of a standard five-element laser, acquired under two different pumping levels: 5 and 4 times above the threshold (Figs. S5F & H). It is worth mentioning that the numerical aperture (NA=0.42) of the 50X objective lens imposes some limitations in capturing the larger diffraction angles in such arrangements.

**Part V. Additional diffraction measurements**

Figures S6A-C demonstrate the normalized emission spectra of another SUSY laser, its constituent single element laser, and the five-element standard laser array, at a pumping power level of approximately 5 times above the threshold of the SUSY laser. Similar to the results presented in the main manuscript, here we also observe that the multiple transverse mode behavior in a standard laser array (Fig. S6C) turns into a single spatial mode operation in the SUSY laser (Fig. S6A). In this regard, the spectral characteristics of the SUSY cavity resembles that of a single element laser where only the $TE_0$ mode is excited (Fig. S6B). However, there are two main differences between the single mode emission from the SUSY array and that of a single element laser. First, the diffraction angle measurements (Figs. S6D-F), performed at two locations along the optical axis (spaced 3 mm away from each other), reveal that the SUSY array in comparison to a single element laser, generates a tighter focus in the far-field. This is an indication that the SUSY array emits in the fundamental mode and its constituent lasers are indeed phase-locked. Second, the SUSY array generates higher output power. This means that the SUSY array delivers higher power to a tighter focus, indicating that the brightness of the SUSY laser indeed scales with a factor larger than $>N = 5$.

**Part VI. Sensitivity analysis**

The single transverse mode operation of the SUSY arrangement shows resilience to first order perturbations and fabrication errors. In order to assess the tolerance of the designed SUSY arrangement to fabrication imperfections, a sensitivity analysis was performed on the superpartner array. Assuming a *perturbing* Hamiltonian $V$, the final system can be described by $H_p^{(2)} = H^{(2)} + \varepsilon V$, where $\varepsilon$ represents the strength of perturbation. To first order, the original eigenvalues $\lambda^{(2)}$ are perturbed according to $\Delta\lambda^{(2)} = \lambda_p^{(2)} - \lambda^{(2)} = \varepsilon\langle\varphi^{(2)}|V|\varphi^{(2)}\rangle$. In the absence of degeneracies, the higher-order terms in the power expansion of $\lambda_p^{(2)}$ can be neglected. Based on the numerically computed eigenmodes $|\varphi^{(2)}\rangle$ (Fig. 2 in the main text), we have directly calculated the parameter $\Delta\lambda^{(2)}$ for an ensemble of different Hamiltonian implementations of $V$. These include both random diagonal and off diagonal perturbations, due to either fabrication errors or nonlinearly induced detunings. The resulting angular histograms (Fig.S7) clearly indicate that the slopes $\langle\varphi^{(2)}|V|\varphi^{(2)}\rangle$ do not exceed $45^0$ and are mostly concentrated in a region of $\pm 15^0$ ($\tan(15^0) = 0.27$) around the ε-axis. The higher-order modes exhibit even less sensitivity to dimension variations, owing to their weaker confinement in the original array. This implies that the applied supersymmetric transformation method can be considered robust against first-order perturbation defects.

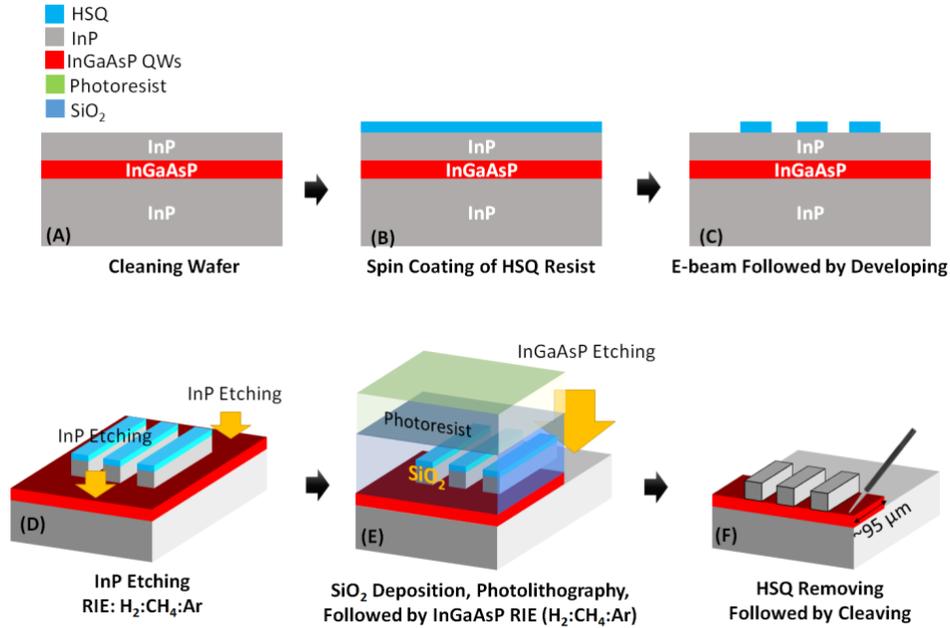

**Fig. S1. Fabrication process.** (**A**) wafer structure, (**B**) spinning HSQ resist, (**C**) electron beam lithography and developing, (**D**) dry etching and transferring patterns into the wafer, (**E**) defining the back facet of the lasers through a four step process, (**F**) removing $SiO_2$ and HSQ by BOE wet-etching, and cleaving the wafer to from the front facet.

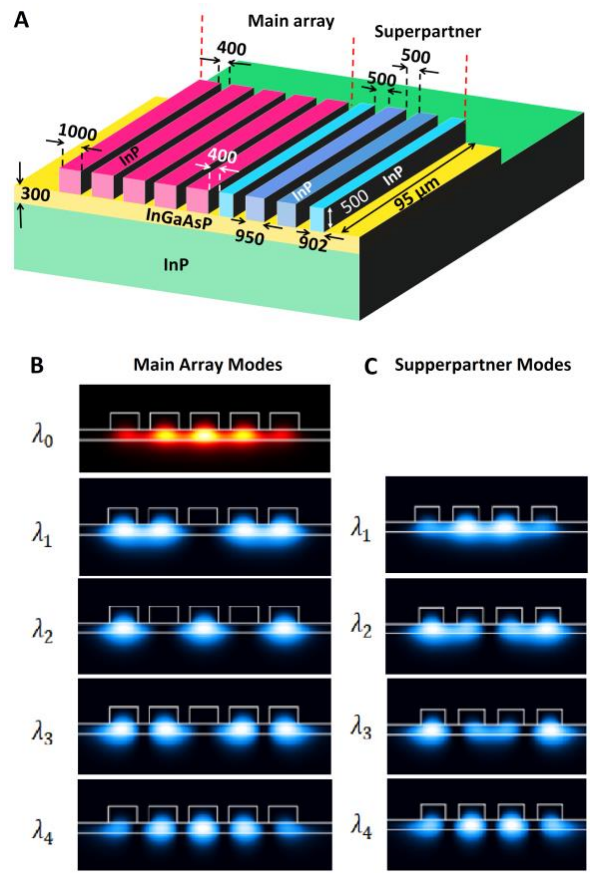

**Fig. S2. Design of the SUSY laser.** (**A**) Schematic representation of the designed SUSY laser. The main array consists of identical waveguides with a 400 nm gap between them. (**B**) The intensity of the modes of the primary array, obtained by FEM simulations. (**C**) The intensity of the modes of the superpartner structure, obtained from FEM simulation. The four higher order eigenvalues, $\lambda_1$ to $\lambda_4$, of the main array are pairwise matched to the eigenvalues of superpartner array.

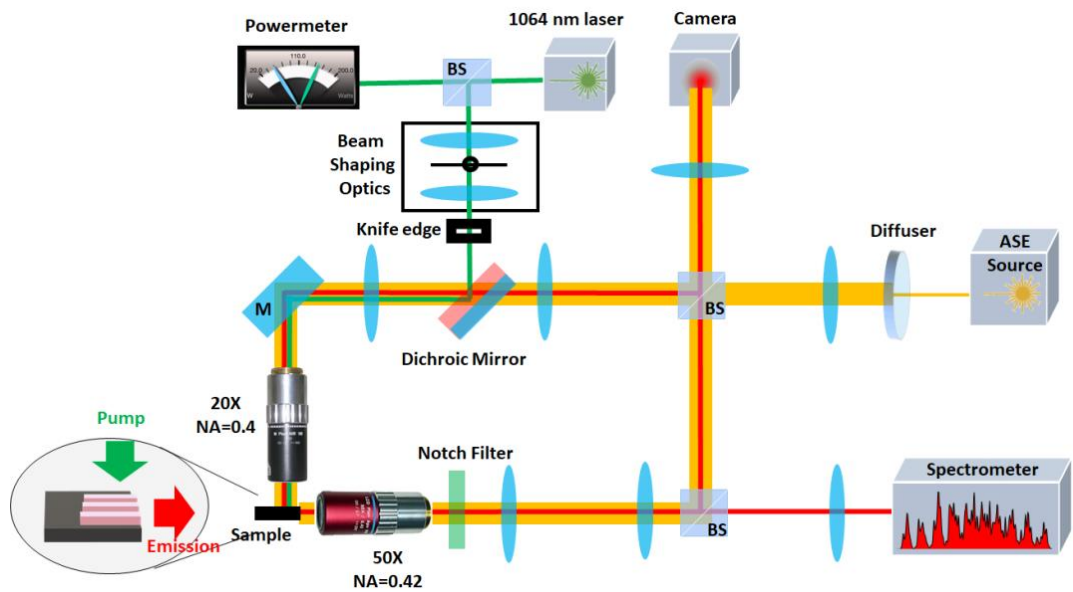

**Fig. S3. Optical measurement setup.** A 1064 nm laser is used as the pump source. After shaping into a proper beam size, it illuminates the sample from the top. A knife edge in the path blocks the pump beam form illuminating the superpartner. Additionally, the 20X objective lens along with an ASE source serve to image the sample from a top view onto an IR camera. The emission from the cleaved facets of the lasers are collected via a 50X objective lens and is directed to a spectrometer and the IR camera to capture the far-field patterns of the samples.

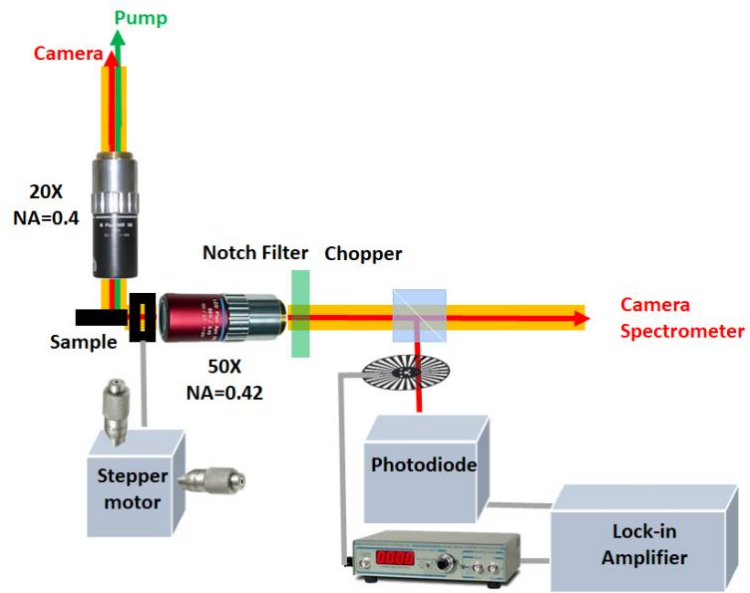

**Fig. S4. Optical setup for diffraction angle measurement.** A rectangular slit is raster scanned along the lasers' slow axis by a stepper motor. In each step, a photodiode in a lock-in detection scheme is used to measure the emitted power passing through the slit.

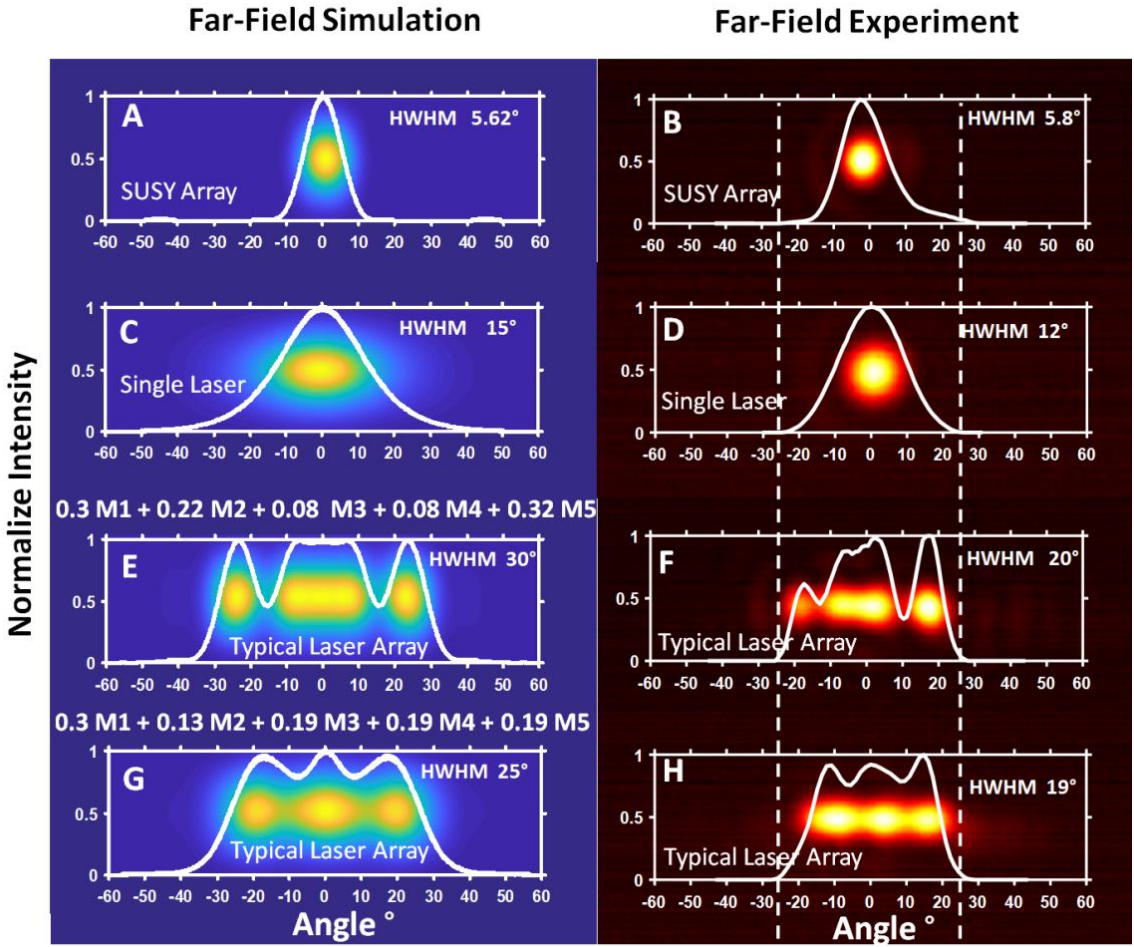

**Fig. S5. Simulated vs. experimental far-field emissions.** Numerically obtained far-field diffraction pattern of (**A**) the SUSY laser array, (**C**) a single element laser, (**E**) the standard laser array at a pump intensity 5 times above threshold, and (**G**) 4 times above threshold. (**B**), (**D**), (**F**), and (**H**) are the corresponding experimental far-field patterns that match to the simulated profiles. The limitation imposed by the numerical aperture of the 50X objective lens in measuring the diffraction angles is highlighted by dashed lines.

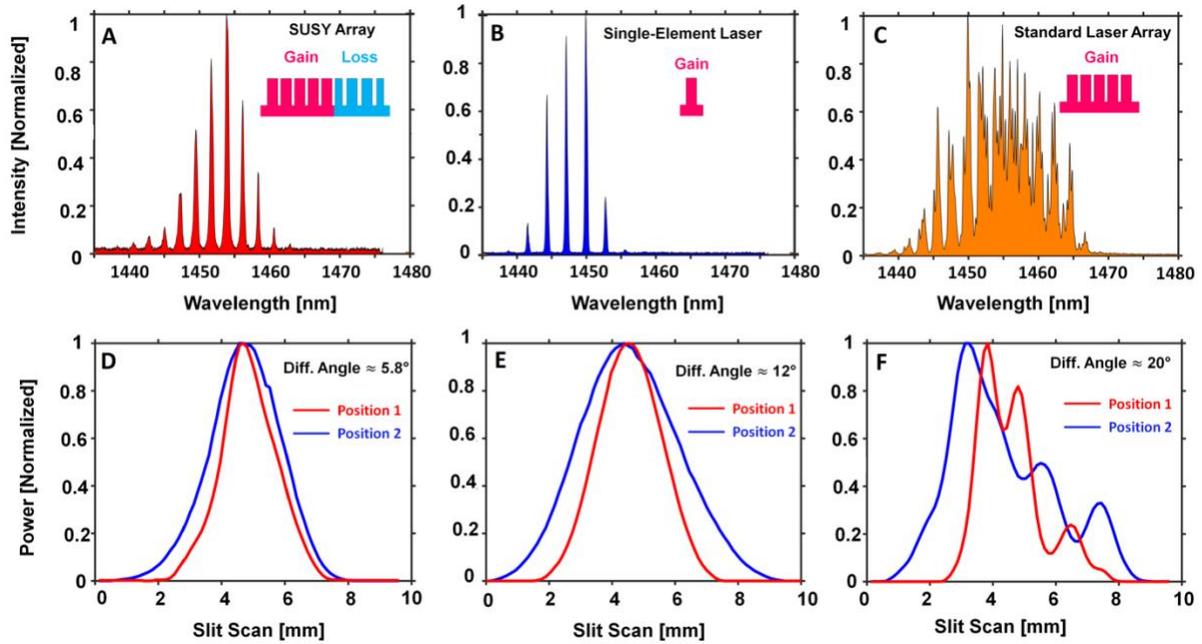

**Fig. S6. Additional spectral and diffraction measurements.** The measured spectra of (**A**) the SUSY laser array, (**B**) a single element laser, and (**C**) a standard laser array at a pump power of about 5 times threshold. (**D-F**) The corresponding diffraction patterns obtained at two different locations along the optical axis, with a separation of about 3 mm. The vertical axis represents the emitted powers of the lasers, measured right after the lens, and the horizontal axis displays the scan range of the rectangular slit along the lasers' slow axis. In these figures, the red and blue lines are linked with the measurements at the two positions along the optical axis. Red curves are obtained at a distance closer to the sample.

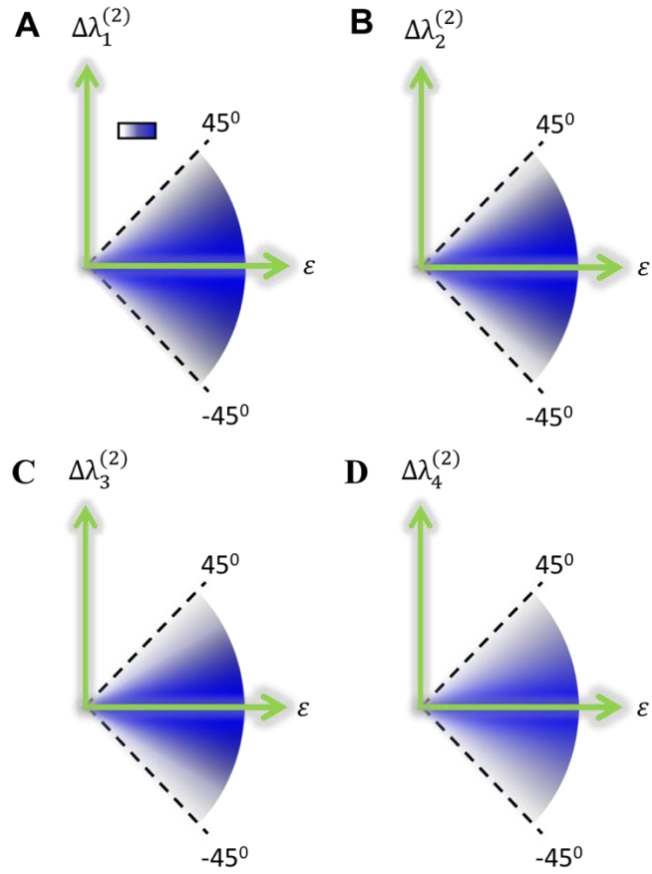

**Fig. S7. First-order perturbation analysis.** Angular histograms of the variations in the normalized propagation constants with respect to perturbations of order $\varepsilon$ in both onsite and off-diagonal elements in the superpartner array for (**A**) mode #1, $\Delta\lambda_1^{(2)}$, (**B**) mode #2 $\Delta\lambda_2^{(2)}$, (**C**) mode #3 $\Delta\lambda_3^{(2)}$, and (**D**) mode #4 $\Delta\lambda_4^{(2)}$. The colorbar inset in (**A**) describes the relative strength $\langle\varphi^{(2)}|V|\varphi^{(2)}\rangle$ of these perturbations (0-100%). The dotted lines indicate the $\varphi = \pm 45^0$ azimuthal lines, and signify a sensitivity threshold.